\documentclass{camera}

\usepackage{epsfig}
\begin{document}

%
\title{A STRONG ELECTROWEAK SECTOR AT FUTURE COLLIDERS}

%
\author{D. Dominici}

%
\organization{Dipartimento di Fisica, Florence University and INFN, \\
 Via Sansone 1, 50019 Sesto. F. (FI), ITALY}
 \maketitle

%
\centerline{\bf Abstract} A brief overview of the production at
future colliders of two new triplets of spin one resonances  from
a strong electroweak breaking is presented. \vspace*{0.2cm}

Fits to the electroweak precision data assuming the Standard Model
(SM) suggest a light Higgs; however the data do not necessarily
exclude possible extensions with a large Higgs mass provided that
its  effect is compensated  by the effect of some new high order
operator or of some new particle \cite{vari}. A recent critical
review of this option can be found in \cite{Peskin:2001rw}. These new
operators or the presence of new particles can give distinctive
signatures at new accelerators like LHC and future linear
colliders. For instance operators of order $p^4$
appearing in the non linear breaking effective lagrangian  can be
detected by studying $WW$ scattering at future colliders with the
sensitivity shown in \cite{chierici}; a review of some
possible signatures concerning new resonances can be found in 
\cite{Barklow:2002su}.
 Among the
models with new vector particles I will present here a brief
overview of
 the phenomenology of
the degenerate BESS model.

The degenerate BESS model (D-BESS)~\cite{dbess} is a realization
of dynamical electroweak symmetry breaking  which predicts the
existence of two new triplets of gauge bosons almost degenerate in
mass ($L^\pm$, $L_3$), ($R^\pm$, $R_3$). The extra parameters  are
a new gauge coupling constant $g''$ and a mass parameter $M$,
related to the scale of the underlying symmetry breaking sector.
In the charged sector the $R^\pm$ fields  are unmixed and
$M_{R^\pm}=M$, while $M_{{L}^\pm}\simeq M (1+x^2)$ where $x=g/g''$
with $g$  the usual $SU(2)_W$ gauge coupling constant. The $L_3$,
$R_3$ masses are given by $M_{L_3}\simeq  M\left(1+ x^2\right),~~
M_{R_3}\simeq M \left(1+ x^2 \tan^2 \theta\right)$ where $\tan
\theta = s_{\theta}/c_{\theta} = g'/g$ and $g'$ is the usual
$U(1)_Y$ gauge coupling constant. These resonances are narrow and
almost degenerate in mass;  this model respects the existing
stringent bounds from electroweak precision data since the $S,T,U$
(or $\epsilon_1, \epsilon_2, \epsilon_3$) parameters vanish at the
leading order in the large $M$ expansion due to an additional
custodial symmetry. Therefore the precision electroweak data only
set loose bounds on the parameter space of the model. 

Future
hadron colliders may be able to discover these new resonances by
their production through quark annihilation and decay in the
lepton channel: $q{\bar q'}\to L^\pm,W^\pm\to (e \nu_e)
\mu\nu_\mu$ and $q{\bar q}\to L_3,R_3,Z,\gamma\to
(e^+e^-)\mu^+\mu^-$. The main backgrounds, left to these channels
after the lepton isolation cuts, are the Drell-Yan processes with
SM gauge bosons exchange in the electron and muon channel. A study
has been performed using Pythia and CMSJET, which performs a
simulation of the energy smearing of CMS detector \cite{redi}.
Results are given in table~\ref{dom:table1} for the sum of the
electron and muon channels for $L=100$~fb$^{-1}$. For the case
$M=$3~TeV the results are given for an integrated luminosity of
500 fb$^{-1}$.
\begin{table}
\caption{Sensitivity to $L_3$ and $R_3$ production at the LHC and CLIC
for $L=$100(500)~fb$^{-1}$ with $M=$1,2(3)~TeV at LHC and
$L=$1000~fb$^{-1}$ at CLIC.}
\begin{tabular}{c c c c c c c}
$g/g''$ & $M$ & $\Gamma_{L_3}$ & $\Gamma_{R_3}$ & $\frac{S}{\sqrt{S+B}}$&
 $S/\sqrt{S+B}$ & $\Delta M$
\\
& (GeV) &(GeV) & (GeV) & LHC ($e,\mu$) & CLIC (had) &  CLIC \\
 \hline  0.1 &
1000 & 0.7 & 0.1 &17.3 & &
\\
0.2 & 1000 & 2.8 & 0.4 & 44.7& &
\\\hline
0.1 & 2000 & 1.4 & 0.2 &3.7& &
\\
0.2 & 2000 & 5.6 & 0.8 & 8.8& &
 \\\hline
0.1 & 3000 & 2.0 & 0.3 &(3.4)& ~62 & 23.20 $\pm$ .06
\\
0.2 & 3000 & 8.2 & 1.2 &(6.6)& 152 & 83.50 $\pm$ .02
\end{tabular}
\label{dom:table1}
\end{table}
The discovery limit at LHC with $L=100$~fb$^{-1}$ is  $M\sim
2$~TeV with $g/g''=0.1$. Beyond discovery, the possibility to
disentangle the double peak structure depends strongly on $g/g''$
and smoothly on the mass~\cite{redi}. A lower energy LC can also
probe this multi-TeV region through the virtual effects in the
cross-sections for $e^+e^-\to {L_3},{R_3},Z,\gamma \to f \bar f $.

\begin{figure}[htbp]
\centerline{
\includegraphics[scale=0.7]{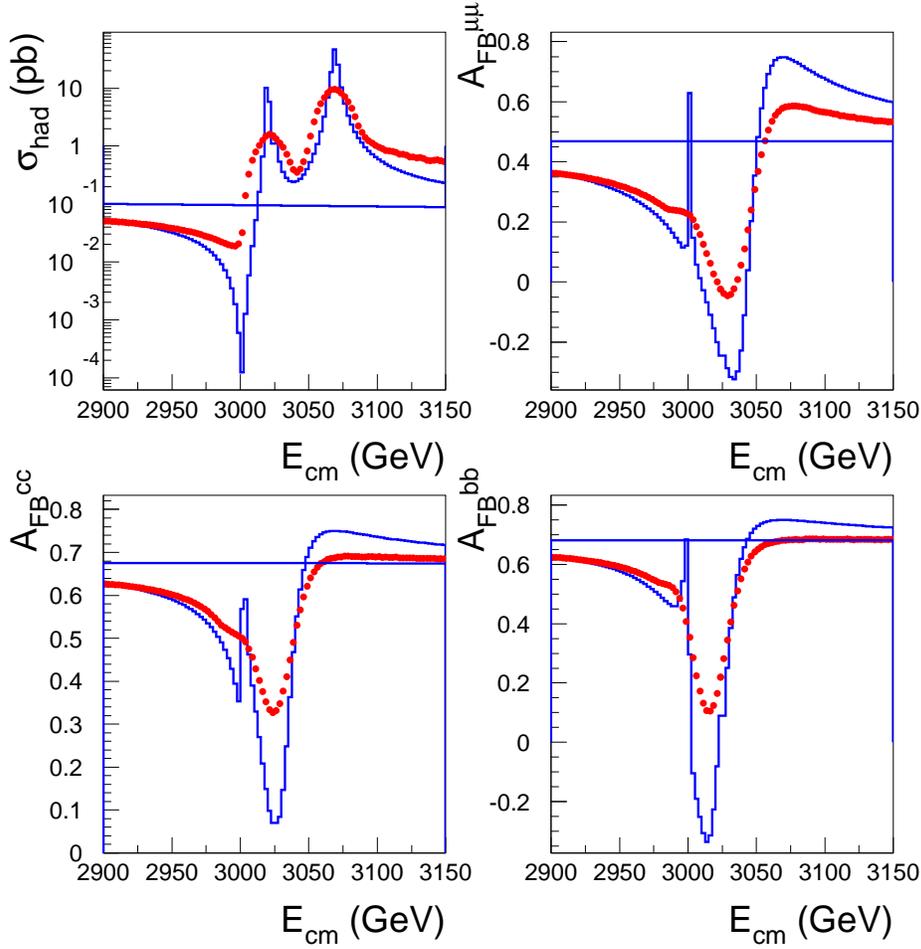}}
\vspace*{0.1cm}
\caption{The hadronic cross section (upper left) and $\mu^+\mu^-$ (upper
right), $b \bar b$ (lower left) and $c \bar c$ (lower right) forward-backward
asymmetries at energies around 3~TeV. The continous lines represent the
predictions for the D-BESS model with $M$ = 3~TeV and $g/g''=0.15$, the flat
lines the SM expectation and the dots the observable D-BESS signal after
accounting for the CLIC.02 luminosity spectrum}
\label{clic}
\end{figure}

Assuming a resonant signal to be seen at the LHC or at a lower LC,
the multi-TeV collider can measure its width, mass and investigate
the existence of an almost degenerate structure~\cite{flab}. The
ability in identifying the model distinctive features has been
studied using the CLIC production cross section and the flavour
dependent forward-backward asymmetries, for different values of
$g/g''$. The CLIC luminosity spectrum has been obtained with a
dedicated beam simulation program  for the nominal parameters at
$\sqrt{s}$ = 3~TeV.  The resulting distributions  for $M$ = 3~TeV
and $g/g''=0.15$ are shown in fig.~\ref{clic}
 for the case of
the CLIC.02 beam parameters
 (a luminosity, $L$=0.40$\times10^{35}$~cm$^{-2}$ s$^{-1}$ and a number of photons radiated per $e^\pm$ in the bunch, $N_{\gamma}$=1.2).

This study has demonstrated that with 1000~fb$^{-1}$ of data, CLIC
will be able to resolve the two narrow resonances for values of
the coupling ratio $g/g'' >$~0.08, corresponding to a mass
splitting $\Delta M$ = 13~GeV for $M=$ 3~TeV, and to determine
$\Delta M$ with a statistical accuracy better than 100~MeV (see
table~\ref{dom:table1}).

%
\def\MPL #1 #2 #3 {Mod. Phys. Lett. {\bf#1}\ (#3) #2}
\def\NPB #1 #2 #3 {Nucl. Phys. {\bf#1}\ (#3) #2}
\def\PLB #1 #2 #3 {Phys. Lett. {\bf#1}\ (#3) #2}
\def\PR #1 #2 #3 {Phys. Rep. {\bf#1}\ (#3) #2}
\def\PRD #1 #2 #3 {Phys. Rev. {\bf#1}\ (#3) #2}
\def\PRL #1 #2 #3 {Phys. Rev. Lett. {\bf#1}\ (#3) #2}
\def\RMP #1 #2 #3 {Rev. Mod. Phys. {\bf#1}\ (#3) #2}
\def\NIM #1 #2 #3 {Nuc. Inst. Meth. {\bf#1}\ (#3) #2}
\def\ZPC #1 #2 #3 {Z. Phys. {\bf#1}\ (#3) #2}
\def\EJPC #1 #2 #3 {E. Phys. J. {\bf#1}\ (#3) #2}
\def\IJMP #1 #2 #3 {Int. J. Mod. Phys. {\bf#1}\ (#3) #2}
\def\JHEP #1 #2 #3 {J. High En. Phys. {\bf#1}\ (#3) #2}
%

%
\end{document}